\documentclass[final,3pt,times,twocolumn,2023]{elsarticle}

\usepackage{amssymb}
\usepackage{multirow}
\PassOptionsToPackage{hyphens}{url}
\usepackage{hyperref}
\usepackage{tikzsymbols}

\journal{Journal of Network and Computer Applications}

\begin{document}
\begin{frontmatter}



\title{Modular zk-Rollup On-Demand\endgraf\textcolor{red}{\rule[+1ex]{\textwidth}{1mm}} \textcolor{red}{Submitted version of paper accepted at Journal of Network and Computer Applications}\endgraf\textcolor{red}{\rule{\textwidth}{0.9mm}}}


\author[ISAE,UPS]{Thomas Lavaur}
\ead{thomas.lavaur@isae-supaero.fr}
\author[ISAE]{Jonathan Detchart}
\author[ISAE]{Jérôme Lacan}
\author[ISAE]{Caroline P. C. Chanel}

\affiliation[ISAE]{organization={ISAE-SUPAERO, University of Toulouse},
            city={Toulouse},
            postcode={31400}, 
            country={France}}
\affiliation[UPS]{organization={University Toulouse III Paul Sabatier},
            city={Toulouse},
            postcode={31062}, 
            country={France}}

\begin{abstract}

The rapid expansion of the use of blockchain-based systems often leads to a choice between customizable private blockchains and more secure, scalable and decentralized but expensive public blockchains. This choice represents the trade-off between privacy and customization at a low cost and security, scalability, and a large user base but at a high cost. In order to improve the scalability of secure public blockchains while enabling privacy and cost reduction, zk-rollups, a layer 2 solution, appear to be a promising avenue. This paper explores the benefits of zk-rollups, including improved privacy, as well as their potential to support transactions designed for specific applications. We propose an innovative design that allows multiple zk-rollups to co-exist on the same smart contracts, simplifying their creation and customization. We then evaluate the first implementation of our system highlighting a low overhead on existing transaction types and on proof generation while strongly decreasing the cost of new transaction types and drastically reducing zk-rollup creation costs.

\end{abstract}




\end{frontmatter}

%
\section{Introduction}
\label{sec:Intro}

First in the field of finance, Bitcoin, a blockchain-based system, has been proposed as a decentralized alternative to traditional currencies \cite{nakamoto2008bitcoin}. Furthermore, during the last decade, the rise of the blockchain has been gaining more and more momentum. More recently, they have even begun being applied in other fields (e.g., the energy sector, \cite{smart_grid}). Their use in these various other fields is primarily motivated by the properties of blockchains such as security, decentralization, immutability, confidentiality and anonymity \cite{abou2019blockchain}.

However, the use of blockchains is not a panacea and their use brings its own limitations and problems. Firstly, we have the limitation in the number of transactions per second (TPS) and the cost of these transactions when the blockchain is highly decentralized. A \emph{trilemma} of balance between scalability, decentralization and security is often mentioned (see \cite{blockchain_trilemma}). Since 2019, the use of rollups has been growing and is becoming increasingly popular as an effective solution to the problems arising with the use of blockchains \cite{blockchain_review}. A rollup is a layer 2 solution for outsourcing the execution of transactions and then verifying them on the blockchain. This allows for the same level of security as on the blockchain \ref{fig:Layers}. 

Optimistic rollups are layer 2 scalability solutions \cite{review_rollup_Tremblay} where all transactions are considered valid, a priori, allowing for a simple and fast system. They rely on fraud proofs for their security, which can bring delays to the finality of transactions. In constrast, zk-rollups can significantly increase the number of TPS on the blockchain while maintaining the same level of security as the underlying blockchain and with instant finality \cite{lavaur2022enabling}. Zero-knowledge protocols, used in zk-rollups, are still new and improvements are being discovered rapidly \cite{Ben_Sasson_blog} in order to reduce their complexity of computation. These solutions provide a significant reduction in transaction costs in exchange for a costly smart contract deployment. Currently, only a few major companies offer the use of permissionless rollups, or validium/zk-rollup services. However, their solutions imply the centralization of zk-rollup ownership which, while not decreasing security, increases the risk of censorship and decreases customization opportunities for users.

In this paper, we claim that zk-rollup technology can technically allow for the replacement of most private blockchains by rollups. They allow for the same properties such as privacy, whitelist management and a higher number of TPS with the same security as the underlying blockchain. Therefore, by using zk-rollups, we can propose the same customizable solutions as those envisaged with the use of a private blockchain while taking advantage of the security, the currency and the connected community of a large and secure public blockchain.

The purpose of this article is to propose this kind of on-demand zk-rollup scheme, where users can easily create their own zk-rollups, without having to design them nor implement and deploy their own smart contract on layer 1 (i.e. the blockchain). As previously argued, this service could replace the creation of a private blockchain where users could create their own zk-rollups, using their own definitions of valid transactions and accessing the properties of a public blockchain. In detail, our proposal facilitates access to the security of the blockchain by giving users control over their zk-rollups, and the ability to make them permissioned and hide transactions from the blockchain if they wish. In this sense, users can control all the important aspects of their system and choose the form in which transactions will be stored on their layer 2. Additionally, users can benefit from the security and community of a public blockchain via its tokens and the validation of their transactions on it. 

This paper is organized as follows. A brief introduction to the blockchain, rollup and zk-rollup concepts is given in Section \ref{sec:StateOfTheArt}. In Section \ref{sec:proposition}, a detailed description of our motivations, the difficulties we had to solve and our proposal are presented. Finally, Section \ref{sec:results} presents an implementation of our proposal, discusses the choices that were made and presents the results related to its evaluation. The paper concludes with a discussion of future work perspectives in Section \ref{sec:conclusion}.

\section{Background}
\label{sec:StateOfTheArt}

\subsection{Blockchains}

\begin{figure*}[th!]
    \centering
    \includegraphics[width=\linewidth]{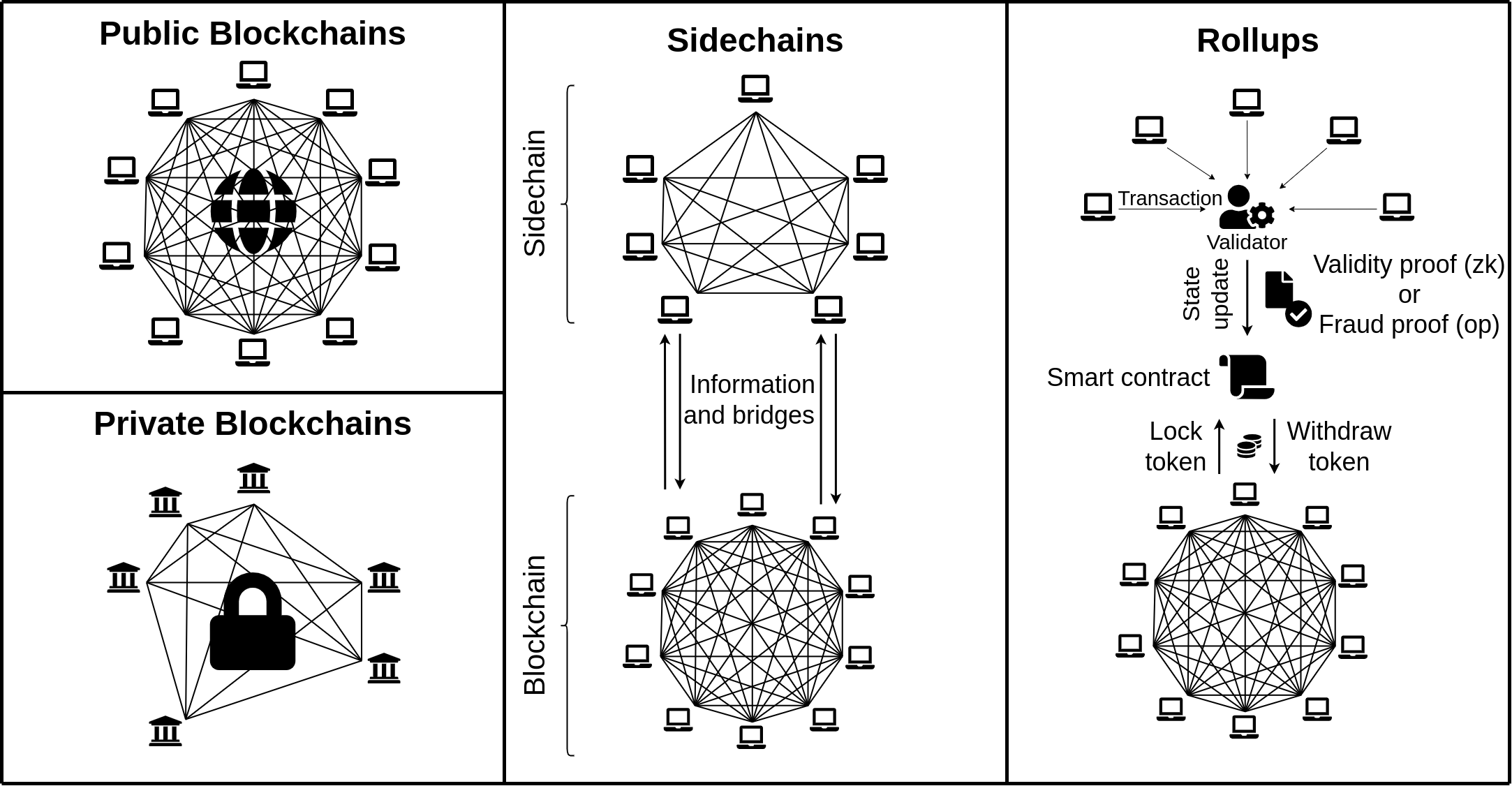}
    \caption{Representation of different Layer 1s and Layer 2s}
    \label{fig:Layers}
\end{figure*}

A blockchain is a distributed and replicated ledger in a network of nodes. The addition of new information in this ledger is regulated by a set of rules common to the nodes that guarantee security. The addition of information is called a transaction and is done linearly without ever erasing the information stored previously. For this, transactions are most often grouped in a structured format and included in what is called a block. A block contains the transactions and all the information necessary for their verification and addition to the blockchain. In order for a transaction to be validated, the sender signs it cryptographically and sends it to the entire network. If a transaction is validated based on the correct signature and the ability of the user to perform the desired action, it is added to the next block. The node in charge of forming the blocks and adding them to the blockchain is designated by the blockchain consensus. All blocks are cryptographically linked each other to form the blockchain. This is often a cryptographic nor game theory protocol that avoids including invalid transactions. Each node can therefore verify whether this new block is valid or not, by checking the transactions within the block. If two valid versions of the blockchain are received at the same time, the longer one is usually considered valid. Depending on the context, it is possible to configure who can write and read on the blockchain, in other words to define whether it is permissionless or permissioned \ref{fig:Layers}.

The first blockchain, Bitcoin, was proposed by Nakamoto in 2009 with the idea of offering a decentralized alternative to traditional currencies \cite{nakamoto2008bitcoin}. In the case of Bitcoin, transactions are limited to the exchange of a valuable asset: the bitcoin. The bitcoin token is created directly on the blockchain as a reward to nodes that add new blocks. This blockchain is public: anyone can become a node, read and write to the blockchain.

In order to expand the capacity of blockchains, currency transactions have been extended to support the execution of programs directly on the blockchain while maintaining security. This is made possible by smart contract technology, which allows for the deployment of code in a first step, which can be executed afterward by all the nodes of the network. When a transaction requests the execution of a function of this code, it is run by the whole network once included in a block to check that the program was processed correctly. Ethereum, whose currency is the ether, allows for the use of smart contracts written in Solidity \cite{buterin2014next}. Its programs can be run in the Ethereum Virtual Machine (EVM) environment \cite{eth_yellow}, an environment deployed on all nodes. Most of the blockchains enabling the use of smart contracts are based on the EVM.

The main properties of blockchains are security, decentralization, immutability, privacy and anonymity \cite{abou2019blockchain}. Security is ensured by the consensus that elects the next node, encouraging it to act honestly. Most consensuses require that a majority of participating nodes are honest so that no invalid transactions are included (usually 50\% or 70\% of participants). If a large number of nodes participates in the consensus, then a high degree of security is ensured. When few nodes participate, there is a risk of centralization meaning that an attack is more likely to occur. Immutability derives from the fact that data can only be added to the register. Anonymity comes from the use of asymmetric keys for authentication and the validation of transactions, allowing for the decorrelation of users' identities from their blockchain accounts (represented by their public key).

In the case of Bitcoin and Ethereum, the nodes responsible for the creation of new blocks are incentivized to be honest through rewards in the form of cryptocurrencies managed by the blockchain. The choice of the transactions to include in the next block is often based on the same principle: users who submit a transaction include a commission fee for the node which includes it in the blockchain. The block creators will therefore choose the transactions that carry the highest commission fees for them. In the case of a private blockchain, or one dedicated to a particular application, the incentive may not be financial. The popularity and high security of Bitcoin and Ethereum generate high transaction costs due to the low number of TPS on these blockchains and the large number of users who are looking to quickly include their transactions. Bitcoin and Ethereum have low throughput in terms of TPS: they only support an average of 7 TPS for Bitcoin and 15 TPS for Ethereum. This is largely insufficient for many applications (for comparison, the Visa network can handle 65,000 TPS \cite{guo2022survey}). To compensate for these difficulties, the use of private blockchains, i.e. where access is restricted, is often preferred. This allows each application to create a unique blockchain whose properties and uses are chosen according to the application's specificities. However, these blockchains involving only a limited number of nodes are often poorly secured, tokens are often not representative of what they are supposed to represent (especially those that are supposed to represent a currency or crypto-currency) and the community is limited by design.

\subsection{Sidechains}

Because of the small number of TPS supported by these blockchains, solutions have been invented to support large data and/or transaction flows \cite{blockchain_review}. Blockchains that can handle many TPS often do so at the expense of security or decentralization. Indeed, in order to get more TPS, it is necessary to be able to process transactions faster and encourage nodes to do so faster. Therefore, nodes should be encouraged to continue processing them for lower fees per transaction and by allocating more of their storage space. Consensus to efficiently handle large numbers of TPS is often less secure and might not scale when the number of nodes gets larger \cite{pbft}.

One potential solution, proposed in 2014, is to create a separate blockchain and implement a bidirectional bridge system, also known as a cross-chain protocol, or as mainchain and sidechain \ref{fig:Layers} when a new blockchain is built on top of an existing one, to perform transactions between them \cite{two_way_peg}. This would enhance customization and give the ability to have custom consensus, rules and security controls through a dedicated blockchain, while being able to communicate with other blockchains. This increases the number of TPS on the overall system and thus can address specific needs. However, this advantage is accompanied by a decrease of decentralization since many users must decide on which blockchain they want to participate in, due to storage, network throughput or computation power limitations, thus decreasing overall security.

\subsection{Rollups}

Rollups are layer 2 solutions \ref{fig:Layers}, meaning that they are based on a pre-existing blockchain in the manner of a sidechain. However, rollups can take several forms and are not necessarily blockchains. The principle of rollups is to externalize the execution of transactions outside the blockchain. This allows for an increase in the number of TPS since they do not need to be executed by the blockchain. At the same time, transaction size can be reduced before being posted on the blockchain, reducing fees (i.e. the signature being no longer necessary for the blockchain and other elements being summarized, see \cite{lavaur2022enabling}). The creation of a rollup is done through the deployment of one or more smart contracts and therefore requires a compatible blockchain. The smart contracts will allow the blockchain's users to send or retrieve information (like tokens) to and from the rollup, and more generally, to interact with it.

The addition of information to the rollup is done by an entity, or a few of them, hence it is often centralized. This entity collects the transactions sent directly to the rollup by users, verifies them and forms a batch. This actor updates the state of the rollup and sends new states and batches of transactions through smart contract calls. The state is then updated on the blockchain as well without having to execute the state transitions that result from the transactions. To help the reader understand, we will call this actor in charge of the creation and verification of transaction batches the validator\footnote{This name differs according to the companies proposing rollups and among others, we can find \emph{validator} from Matter Labs and Aztec, \emph{operator} from Starkware and Consensys, \emph{sequencer} and \emph{aggregator} from Polygon, \emph{relayer} from Loopring, \emph{relayer} and \emph{sequencer} from Scroll, etc.}. The state of the rollup is represented via one or more Merkle Trees \cite{merkle_tree} and stored on the smart contract of the rollup through its root. This method is similar to the state of Ethereum accounts where the root of the Merkle Patricia tree is stored in the header of each block \cite{eth_yellow}. In both cases, the trees can be reconstructed by consulting the blockchain history of transactions for blockchain state, or transaction summaries in the case of rollups.

However, to inherit the security of the underlying blockchain, the blockchain must verify the correct execution of the transactions. To our knowledge, it is possible to classify all rollups into two distinct categories according to this verification characteristic: optimistic rollups, where the transactions are all considered valid but can be challenged during a dispute period, and zk-rollups which cryptographically prove the validity of all transactions.

\subsubsection{Optimistic rollups}

In the case of optimistic rollups, all transactions are considered valid, a priori. They can be disputed (i.e. their validity can be questioned) by any user present on the blockchain, by making a request to the smart contract. The validator must then provide the complete transaction to the blockchain in order to verify the signature and produce a Merkle proof attesting to the availability of the tokens or the necessary rights. If no users dispute a transaction, then it is considered valid after a fixed period of time called the dispute period. As the validation of an optimistic rollup transaction on the blockchain is more costly than a native transaction, the system encourages the denunciation of invalid transactions by users and proper publication by the validator, and discourages false claims of invalid transactions \cite{review_rollup_Tremblay}.

\subsubsection{Zk-rollups}

In contrast, zk-rollups prove the validity of all transactions using zero-knowledge cryptographic proofs. The zero-knowledge property allows the blockchain to check the validity of the rollup state update without having to read the transactions while making sure that they are correct and that they are indeed those posted \cite{lavaur2022enabling}. The transactions are therefore immediately validated by the blockchain and a dispute phase is thus unnecessary. Moreover, the validator cannot include invalid transactions. A dishonest validator can only partially censor the transactions but their execution can be forced using the blockchain.

\subsubsection{Validium}

Since the transactions are validated without being read, it is possible to avoid posting the transactions on the blockchain. In this case, these are referred to as validiums \cite{lavaur2022enabling} and not zk-rollups. To ensure security for validiums, the data must be accessible outside the blockchain guaranteeing access to funds or users' information in case of a untrustworthy validator. A database is generally set up by known actors who sign blocks of transactions attesting to the receipt of transaction batches and promising to make them available. It is possible to build privacy on validiums by restricting access to the database to authorized parties while preserving the same security.

\section{Proposition}
\label{sec:proposition}

\subsection{Motivations}

As presented in Section \ref{sec:StateOfTheArt}, the use of zk-rollups is an excellent solution to reduce per-transaction costs and increase the maximum number of TPS on a public blockchain without compromising on security. In parallel, many companies and multiple applications are considering the use of private blockchains to gain more privacy, to obtain a secure private network or to reduce costs. However, the low number of participants in such private blockchains leads, in most cases, to a weakening of security. Moreover, some consensuses, like PBFT, are regularly used on private blockchains and do not allow scaling up when the number of nodes increases \cite{complexity_consensus}.

We claim that zk-rollups are an efficient solution to meet the needs of private blockchains and the security of a public blockchain with numerous participants. Thanks to our proposition, they allow users to take advantage of pre-established communities, pre-established cryptocurrencies and pre-audited security while offering the flexibility of private blockchains designed for specific purposes. The zero-knowledge proof circuit (used to generate proofs) defines the validity of transactions and can be modified from one zk-rollup to another, making it possible to support the proof of several operations dedicated to the rollup, whether they are native to the blockchain or not. To support our consideration of zk-rollups, the Ethereum blockchain itself is moving towards becoming a highly secure blockchain relying heavily on rollups for scalability \cite{buterin_roadmap}.

Unfortunately, zk-rollups are deployed through multiple smart contracts, which generates a significant financial cost. Setup also requires significant expertise since one must: (i) both develop and audit a zero-knowledge circuit that will correctly prove the validity of authorized rollup operations, (ii) establish safe implementation of the smart contracts, and (iii) set up a central server that will play the role of validator. The use of an already deployed zk-rollup does not replace a private blockchain since the few zk-rollups available on the Ethereum blockchain are controlled by a handful of companies that control their validators. Such exclusive control does not allow for the addition of dedicated operations or customization, like whitelist management, and furthermore it ensures visibility on all transactions to the company validator. One solution put forward by different companies is to extend these services providing privacy and customization through layer 3s \cite{l3_starkware} built on top of their own rollup. However, if the layer 2 validator becomes untrustworthy, it is not clear how this would affect the different layer 3s. Furthermore, their validator would still have visibility on all layer 3 transactions.

Moreover, in the case where several zk-rollups are deployed, a user who wants to transfer their token or information from one zk-rollup to another should check the security of all smart contracts which requires knowledge of all addresses they interact with. These steps demand a lot of effort from the user and generates a high cost due to the multiple transactions needed on the blockchain (i.e. the mainnet) to perform this transfer. 

\subsubsection{Modular zk-rollup on-demand}

In order to improve the privacy, customization and cost, and reduce the difficulty of deployment of a zk-rollup, we propose allowing several zk-rollups to co-exist on the same smart contract, in an independent way, without decreasing the security of a single zk-rollup. To do this, we propose including a group ID (or zk-rollup ID) system directly into the smart contracts as already evoked in \cite{lavaur2022enabling}. Our proposition replaces the consecutive deployment of several zk-rollups by a simple call to the smart contracts via a function enabling the creation of a group. This drastically reduces the cost of deployments following the initial deployment. The functions of the smart contracts are shared by the different groups but it is possible to choose a specific smart contract for proof checking in order to use different circuits (i.e. different valid state transition) and proof systems and thus allow for the use of different operations from one group to another. Each group can choose to use the same smart contract as another group, or create its own verification system.

At the same time, we propose the addition of two key parameters specific to each group. The first one defines whether it is a validium or a zk-rollup (i.e. if the validator has to publish transaction data on the blockchain or not) and if the smart contract has to check the signatures of the actors attesting to their availability. This first parameter makes it possible to determine the level of exposure of the transactions on the blockchain without losing security. The second parameter lets the user choose between a permissionless group, open to any user, or a permissioned group, where access and transfer rights are limited to users on a whitelist. Each group can choose its validator and independently store its Merkle tree root representing the rollup state. 

\begin{figure}[th!]
    \centering
    \includegraphics[width=\linewidth]{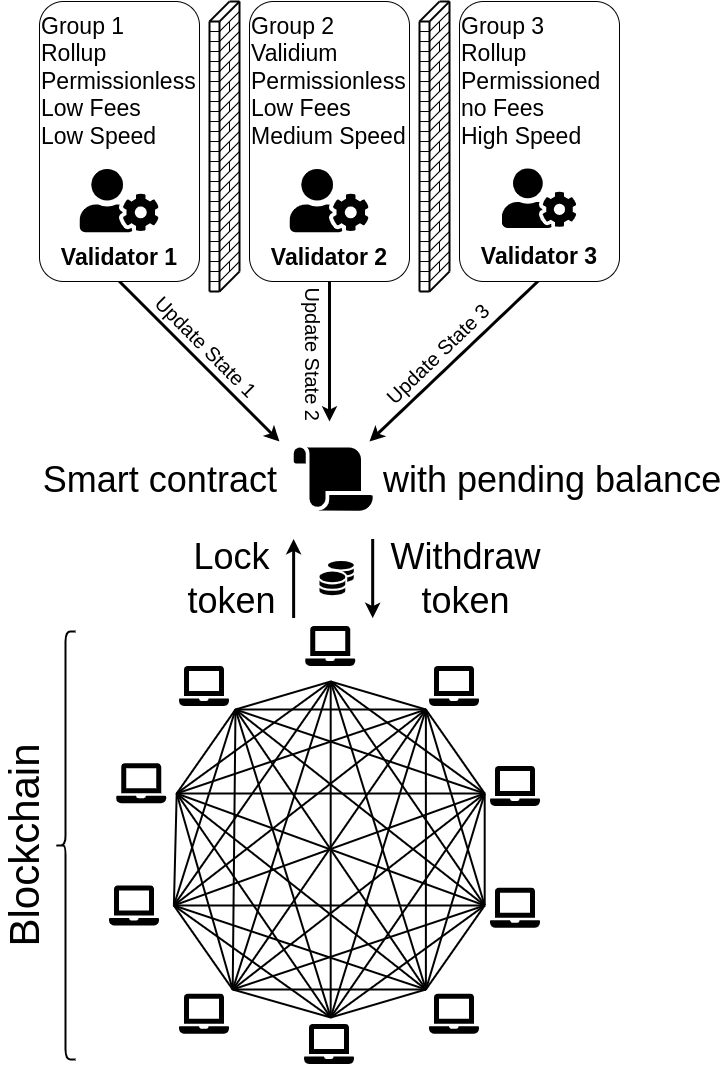}
    \caption{Several zk-rollups on the same smart contract.}
    \label{fig:proposition}
\end{figure}

We advocate that this proposition solves privacy issues while democratizing easy access to zk-rollups for wider adoption. Interestingly, it enables the creation of multiple zk-rollups on the same smart contracts. It can be very interesting even if they are all public and permissionless. Note that most of the current zk-rollup validators are remunerated with fees because the validators also have expenses related to the publication of data on the blockchain, calls to smart contracts and the computation of zero-knowledge proof. With this proposition, we can deploy different zk-rollups with higher or lower fees and with validators who call the blockchain at different frequencies allowing for more or less quick finality on the blockchain. Our solution is summarized in Figure \ref{fig:proposition}.

It is worth recalling that users may later want to change groups or effect an inter-group transaction. To do so, we propose adding a new transaction type that can be interpreted by smart contracts. The main idea is to easily allow users to send/receive information or funds from one group to another without having them to return to the user's address (or require additional transactions on the mainnet). The information remains locked on the smart contract during the entire process. This is particularly important because it would provide the first opportunity for applications to be deployed across multiple zk-rollups, segmented by usage. In particular because depending on the application, the validator may not want to charge fees if their interest or the zk-rollup's interest is not financial.

Thus, with this proposition, it is possible to create a dedicated zk-rollup at low cost, easily and on-demand, managed by the user who creates the group. Such a zk-rollup can be customized to meet privacy and management needs. With our proposal, it is possible to: (i) obtain the properties of the private blockchains we are interested in while also maintaining the security of a public blockchain, (ii) exchange information from one group to another efficiently and without difficulty as only one transaction, at a reduced cost, is required.

\section{Evaluation Methodology}
 \label{sec:results}

In order to demonstrate the feasibility of our proposition and to measure its impact on the performance of a zk-rollup, we modified a pre-existing zk-rollup. We chose the Ethereum blockchain since it currently hosts the majority of zk-rollups and because it has the largest number of users. At the same time, it is the most secure blockchain among those that can interpret smart contracts. Among all the zk-rollups we could rely on, we chose to use zkSync v1\cite{zksync} mainly because it is open source. In the following, we present zkSync v1 in detail.

\subsection{The zkSync' zk-rollup}
 
ZkSync v1 is a zk-rollup developed by Matter Labs already established on the Ethereum mainnet and its security has already been tested \footnote{see \url{https://docs.zksync.io/updates/security-audits/}}. Their code is written in Rust, already flexible and well segmented. As of January 25\textsuperscript{th}, 2023, Matter Labs is a key player and its zk-rollup has the lowest transaction fees on the Ethereum blockchain after Loopring and have the third highest locked-in value behind dYdX and Loopring (other zk-rollups).

The zkSync zk-rollup is operated by several actors. The validator collects formatted and signed transactions from users through a server. It also manages a database in which it stores all information about the status of the rollups. One or more provers connect and retrieve the list of zero-knowledge proof tasks from this database and generate the aforementioned proofs. One or more governors are responsible for the security of the smarts contracts and name the validators. After a publication delay and an agreement, they are able to update the smarts contracts. Several other actors are responsible for specific tasks on the zk-rollups, specifically, counterattacking the potential presence of an untrustworthy validator, enabling a complete customization of each role.

ZkSync v1 is based on the PLONK proof system \cite{Plonk} which requires a trusted setup to generate keys (keys can be use for any circuit). These keys can be generated via the powers of tau ceremony \cite{powersoftau} which is a multiparty protocol. By using this protocol, it becomes possible to generate false proofs only if the totality of the actors who created the key are untrustworthy and coordinate their actions. This protocol ensures strong security.

\begin{figure*}[!bt]
    \centering
    \includegraphics[width=0.7\linewidth]{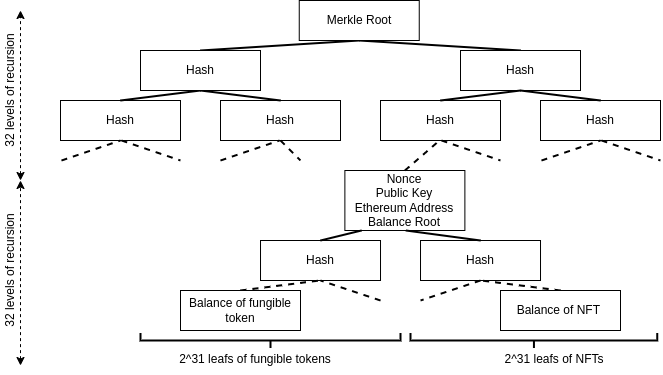}
    \caption{Illustration of accounts representation used in zkSync v1.}
    \label{fig:Merkle_Tree}
\end{figure*}

ZkSync's zk-rollup is based on two different circuits. The first one proves that a batch of transactions, called a block, has been executed in order to update the rollup state. It takes the hash of the block as its public input and outputs the new state of the rollup. The second circuit is a recursive circuit that proves that several proofs from the first circuit were correctly and successfully verified. Thus, the verification is done indirectly as the blockchain will then verify the proof that one or more proofs are correct and that consequently the transactions of several blocks were properly executed. For our proposal, the second circuit did not need to be modified. We therefore reused the same trusted setup employed by zkSync to prove our new circuit (presented later).

The accounts state in zkSync v1 is represented by a single Merkle tree \cite{merkle_tree} of depth 64. The first 32 levels store up to $2^{32}$ different accounts. The leaves store the nonce, the Ethereum address, the zkSync public key, and the root of an account-specific Merkle tree of depth 32 that allows up to $2^{32}$ different token-related amounts to be stored. The first $2^{31}$ are reserved for fungible tokens, i.e. ETH and ERC20. The next $2^{31}$ are for NFTs (Non Fungible Tokens). This structure is illustrated in fig~\ref{fig:Merkle_Tree}. Thanks to this, the transactions published on the blockchain can be summarized by replacing the Ethereum addresses and the tokens used by their respective indexes in the Merkle tree. The update of the accounts state is a three-step process. First the validator must commit one or more blocks that they want to execute and transmit the summary of the transactions to the blockchain. Afterwards, they send the aggregated proof of one or more transaction blocks. The smart contract checks the proof and verifies that the public input used corresponds to the block that was committed (through the hash of the block). In the third step, the validator triggers the execution of one or more blocks so that the on-chain operations can be executed (like withdrawals for example).

Another important component is smart contracts' pending balance on the smart contracts which is in fact the amounts of money that users have locked in the smart contract but which is not integrated into the zk-rollup state. This is also where users' funds are kept inside smart contracts before they are withdrawn. In order to submit batches of transactions, zkSync v1 splits blocks into chunks which are packets of 10 bytes. The current version of zkSync v1 supports 11 types of operations \cite{zksync} which are summarized in table \ref{table:operations}. 

\begin{table*}[h!]
    \centering
    \resizebox{\textwidth}{!}{%
    \begin{tabular}{|c|c|c|c|}
        \hline
        Transaction Type & Size in Chunk & Size in Bytes & Effects \\
        \hline
        \hline
        Deposit	 & 6 & 45 & Request to deposit ETH or ERC20 on the rollup.\\
        \hline
        TransferToNew &	6 & 40 & Funds transfer to a new zkSync address inside the rollup.\\
        \hline
        Withdraw & 6 & 47 & Withdrawal of ETH or ERC20 to the Pending Balance from rollup state.\\
        \hline
        Transfer & 2 & 20 & Funds transfer to a pre-existing address inside the rollup. \\
        \hline
        FullExit & 11 & 85 & Request from the blockchain to withdraw all funds to the pending balance.\\
        \hline
        ChangePubKey & 6 & 49 & Change or set zkSync address (public key).\\
        \hline
        ForcedExit & 6 & 51 & Request from the blockchain to withdraw locked account funds.\\
        \hline
        MintNFT & 5 & 47 & Minting of an NFT inside the rollup.\\
        \hline
        WithdrawNFT & 10 & 95 & Withdrawal of an NFT to the Pending Balance.\\
        \hline
        Swap & 5 & 46 & Swap of two tokens between two accounts inside the rollup.\\
        \hline
    \end{tabular}
    }
    \caption{List of zkSync v1 supported operations}
    \label{table:operations}
\end{table*}

\subsection{Modifications}

To implement our proposition, we added a group field of 16 bits to all transactions natively present on zkSync v1, enabling the creation of a maximum of $2^{16}$ groups on the smart contract. Thus, users sign the group ID indirectly when they sign their transactions. This ID is not inserted into the blockchain when committing a block since the verification of the correct group state update is verified via zero-knowledge proofs. To make sure that the right validator handles the right group's transactions, we forced the validator, who provides the proof, to give their group ID as a public input. By replacing the block hash with the sum of the block hash and the validator group ID inside the public input of the first verification circuit, there is no longer any need to modify the second proof of the aggregation circuit.

During the first proof, the validator also provides their group ID as a secret input and can therefore manipulate it. However, the proof subtracts this group ID from the public input to obtain the hash of the block. Two things are checked in the proof to ensure that the validator does not manipulate the group it processes. The first is that the group ID indicated by the validator allows the reconstruction of the transaction that was signed for each user resulting in correct transaction signature. And the second is that the hash of all the transactions gives the same hash as that of the public entry from which the group ID was subtracted. This first proof ensures that the entire block and all of its transactions are linked to the same group. To bind the validator to a specific group and not let them add another group's block to their own group, we have linked the validators' Ethereum addresses to their group directly via a mapping on the main smart contract. 
Thus, the blockchain rebuilds the public entry itself from the block hash specified during the commit and from the group deduced from the Ethereum address of the user calling the smart contract (presumably the validator) when verifying the proof.

Concretely, we have modified the smart contract to include a group structure and added several fields such as a group identifier (or zk-rollup ID), a Boolean indicating whether the group is permissioned or not, and a mapping to manage a whitelist. Indeed, we propose only allowing users who are not on the whitelist to withdraw funds from the rollup, which ensures continuous management without giving the validator the ability to freeze the funds of removed users. 
In order to enable permissioned zk-rollups, we have added a parameter indicating whether the account is authorized or not into the database and a function to add or remove an Ethereum address from the whitelist into the smart contract. We have not dealt with the case of validium in our implementation leaving it for future work.

Finally, in order to facilitate transactions between two distinct groups, we have created two new types of transactions: ChangeGroup and FullChangeGroup, inspired by Withdraw and FullExit (see Table \ref{table:operations}). Both of these transactions enable users to exit the zk-rollup and retrieve their funds on the blockchain (in the first case, through the validator, and in the second, through the blockchain). These new transaction types enable users to do the same things as Withdraw and FullExit directly to an existing group. Thus, a user who wants to make an inter-group transaction can perform a single transaction on their original group instead of having to withdraw and then call the smart contract to retrieve their funds from the pending balance before finally depositing them on the destination group. This only requires knowledge of the destination group's identifier, reducing the risk of human error compared to several smart contracts. These two new transaction types have exactly the same structure as Withdraw and FullExit but include the original group and destination group. In order to be interpreted, modifications have been made to the block execution and FullExit functions.

\subsection{Setup and metrics}

To measure the performance of our implementation, i.e. the impact of the addition of a group structure and the new transaction types on users, we focused primarily on the measurement of gas costs before and after the modification of several operations. The gas is a unit representing the work that a node will have to do to execute a smart contract, and therefore does not fluctuate with the price of the token on the blockchain. We have used the Hardhat local network to realize the gas measurements of transactions and the main functions of the smart contracts. Additionally, we compared the performance of our new operations (i.e. their gas costs) to that of a user wanting to change zk-rollups in a case where two zkSync v1 rollups would be deployed at different addresses and on different smart contracts.

We also measured the number of gates (which is used to compare complexity) in the first zero-knowledge circuit with its execution time. To compute the proofs, we used a computer with an Intel Xeon Platinum 8164 CPU and 400GB of RAM. For the three phases of the rollup state update, we measured the three functions: the block commit, the block proof and the block execution individually by varying both the size of the blocks (i.e. 26, 78, 182 and 390 chunks or 260, 780, 1820 and 3900 bytes) and the number of proofs that are aggregated (i.e. 1, 4 and 8) - and thus the number of blocks executed and committed at the same time). 

We have separated the costs of each transaction type into six categories: commit base costs, prove base costs, execution base costs, commit extra costs, execution extra costs, and external costs. We have calculated the basic costs of a transaction by relating the size of the operation to the total cost of a fixed number of blocks of a given size. For instance, if a transfer takes up 2\% of the space of a block, its basic cost is 2\% of the cost of a block. For the basic cost of a block, we measured the cost of a block filled only with transactions of type Transfer because they do not trigger any on-chain operations during execution. Note that, block commit overheads are related to the collection of on-chain operations when committing a block. Thus, an operation that will have to be executed on the blockchain will need additional processing on the smart contract as soon as the commit operation is performed, generating an additional cost compared to an internal rollup operation. The same principle applies to the extra costs of the execution function. Finally, the external costs are costs payed by the user directly on-chain, for example, to make a deposit by calling a smart contract.

Finally, our measurements have been made in a harmonized way between commit, proof and execution: when we aggregated 8 blocks, we also committed and executed 8 blocks when calling the functions. It is important to note that these choices are flexible and that a validator can choose an arbitrary number of aggregated blocks to commit, a different number to prove and yet another to execute.

\subsection{Results}

The addition of the two new operation types, the inclusion of the group in the transactions and the modification of the public input create almost no overhead for the prover for the first circuit, the second circuit remaining unchanged. The size of the first circuit only increases from 0.18\% for the smallest blocks to 0.32\% for the largest blocks, and the difference in proof time is not significant. These results are summarized in Table \ref{tab:circuit1}.

\begin{table}[!th]
    \centering
    \resizebox{\columnwidth}{!}{%
    \begin{tabular}{|c|c|c|c|c|}
        \hline
        Block Chunk Size & \multicolumn{2}{|c|}{zkSync} & \multicolumn{2}{|c|}{Our Proposition} \\   
        \hline
        \hline
        26 & 8,526,701c & 71s & 8,542,124c & 71s \\
        \hline
        78 & 16,908,690c & 142s & 16,952,713c & 144s \\
        \hline
        182 & 33,672,019c & 289s & 33,773,242c & 289s \\
        \hline
        390 & 67,185,536c &	588s	& 67,401,159c &	588s \\
        \hline
    \end{tabular}
    }
    \caption{First circuit comparison (c mean constraints, s seconds).}
    \label{tab:circuit1}
\end{table}

Our measurements show that our proposal has a very small negative impact on the execution cost of an individual transaction when block size is the largest and the number of aggregated proofs is the highest. In fact, in these conditions, the cost of a deposit is only increased by 3\% for ERC20 and 2\% for ETH, while the rest of the transactions only see their costs increase by less than 1\%. However, the overall benefits of our new operations are rather impressive as the ChangeGroup operation reduces gas consumption by more than 49\% for ETH and more than 61\% for ERC20. The gains are even more marked when we consider that this operation allows the user to reduce the number of actions required.

\begin{figure*}[h!]
    \centering
    \vspace*{-30mm}
    \centering
    \includegraphics[width=0.95\linewidth]{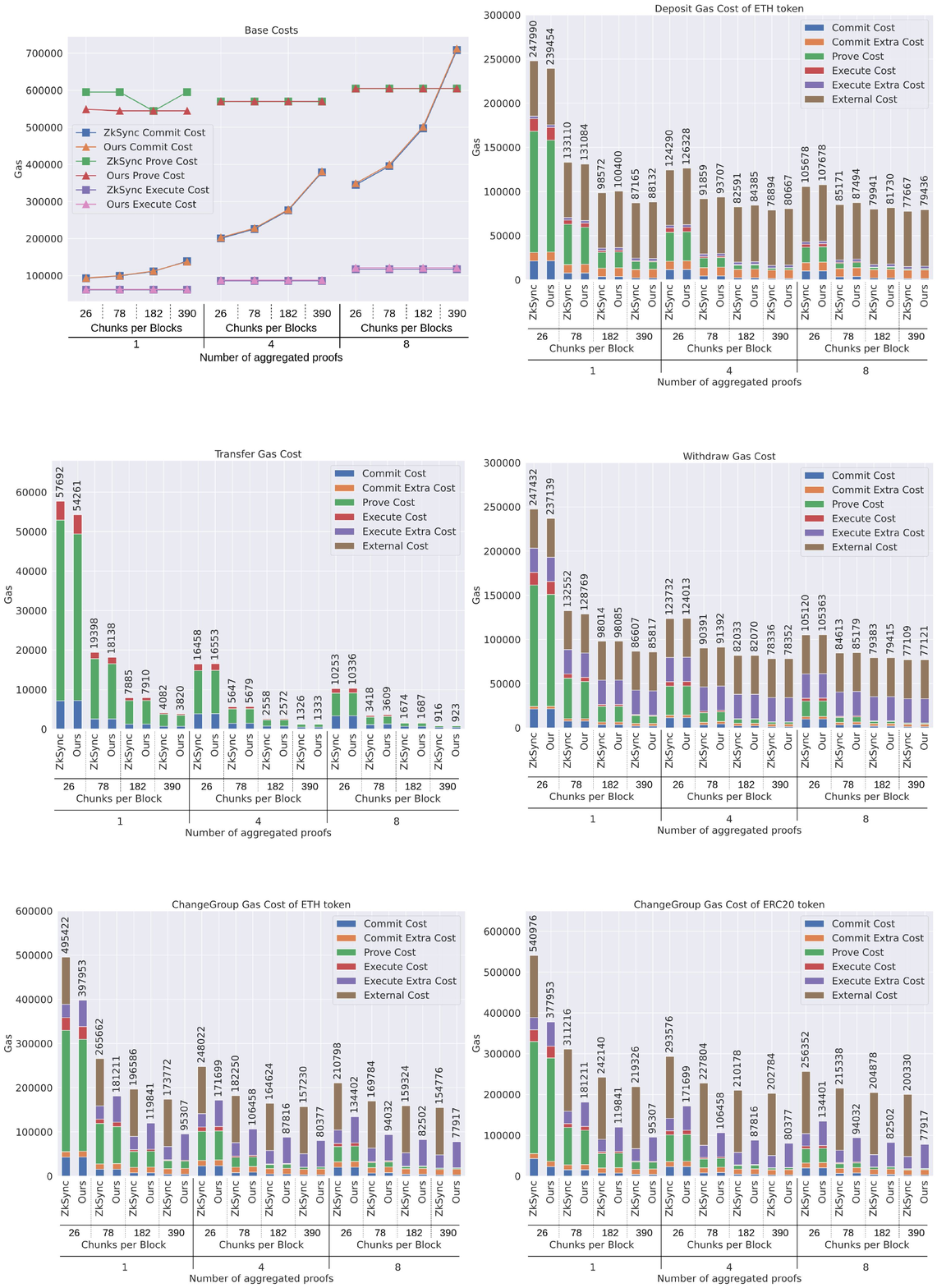}
    \vspace{-8mm}
    \caption{Costs of the main transaction types.}
    \label{fig:Costs}
\end{figure*}

The base costs and the costs of the most used transactions, described in the six categories above, are shown in Figure \ref{fig:Costs}. The other costs and respective resuts tables are available with our implementation in our github repository\footnote{\url{https://github.com/thomaslavaur/Modular-zk-Rollup-On-Demand}}. It is important to highlight that the cost of creating a new zk-rollup, in addition to being simplified and customizable, is drastically reduced with our proposal. In fact, during the first deployment of the smart contracts, our proposal leads to an additional cost of about 4\%, going from 22,106,772 gas to 22,904,219 gas. However, when we compare the cost of redeploying zkSync v1 with the cost of creating a group with our proposal, costs are reduced more than 99\% from 22,106,772 gas (zkSync v1) to 184,258 gas (ours). 

\section{Conclusion and future work perspectives}
\label{sec:conclusion}

In this article, we propose an innovative design to provide an on-demand zk-rollup creation service. It is customizable and can be used as a solution for applications usually requiring a private blockchain while looking for strong security and/or a pre-established community. We propose allowing several zk-rollups to co-exist as groups on the same smart contracts. At the same time, we propose a new type of transaction that allows simple and low-cost transfers from one group to another. We implemented and measured our propositions and concluded that they did not significantly impact the proofs' performances or the costs of existing operations and drastically reduced the costs of new operations introduced and the costs of deployment of the smart contracts. The easy creation of zk-rollups through smart contracts and via group partitioning is highlighted by the reduction in the cost of setting up a zk-rollup: by more than 99\%.

However, until now, our implementation does not enable the creation of a group that would be a validium. This implementation may be necessary and is left for future work.
Moreover, we have developed our implementation over zkSync v1, and we believe it would be interesting in the near future to apply it to a zkVM-compatible zk-rollup, which can prove the execution of smart contracts inside the zk-rollup. ZkSync's zkEVM has recently been made public and open source (unfortunately, just after the end of our implementation).

Finally, many other questions deserve to be explored to improve the speed of finality of zk-rollups. One example is the study of the possibility of pooling the construction of the aggregate proofs of several groups at the same time. For instance, a group could commit and execute a single block but only pay for one-eighth of the aggregated proof of 8 blocks, sharing the cost among 8 groups. We believe such an aggregation could reduce costs even more.\\
\newline\\
This research did not receive any specific grant from funding agencies in the public, commercial, or not-for-profit sectors.

\newpage

\bibliographystyle{elsarticle-num} 
\bibliography{biblio}

\end{document}